\documentstyle[Emulapj]{article}

\begin{document}

\lefthead{MAGNETOHYDRODYNAMIC ORIGIN OF JETS}
\righthead{LOVELACE ET AL.}

\accepted{}

\title{Magnetohydrodynamic Origin of Jets
from Accretion Disks}

\medskip

\author{R.V.E. Lovelace}
\affil{Department of Astronomy,
Cornell University, Ithaca, NY 14853-6801;
rvl1@cornell.edu }
\author{G.V. Ustyugova}
\affil{Keldysh Institute of
Applied Mathematics, Russian Academy
of Sciences, Moscow, Russia, 125047,
ustyugg@spp.Keldysh.ru}
\author{A.V. Koldoba}
\affil{Institute of
Mathematical Modelling,
Russian Academy of Sciences, Moscow,
Russia, 125047}

\slugcomment{For IAU Proceedings No. 194,
Activity in Galaxies and Related Phenomena, August 17-21,
1998, Byurakan, Armenia}

\begin{abstract}
A review is made of recent
magnetohydrodynamic
(MHD) theory and simulations
of origin of jets from accretion
disks.
    Many compact astrophysical objects
emit powerful, highly-collimated,
oppositely directed jets. Included are
the extra galactic radio jets
of active galaxies and
quasars, and old
compact stars in binaries,
 and emission line jets in young stellar
objects.
   It is widely thought that these
different jets arise from rotating,
conducting accretion disks
threaded by an ordered magnetic field.
   The twisting of the ${\bf B}$ field
by  the rotation of the disk drives
the jets by magnetically extracting
matter, angular momentum, and energy from
the accretion  disk.
   Two main regimes have been discussed
theoretically, hydromagnetic winds
which have a significant mass flux,
and Poynting flux jets where the
mass flux is negligible.
     Over the past
several years,  exciting new developments
on models of jets
have come from
progress in  MHD simulations which
now allow the study of the origin -
the acceleration and collimation - of
jets from accretion disks.
   Simulation studies  in the
hydromagnetic wind regime
indicate that the outflows
are accelerated close to their
region of origin  whereas the
collimation occurs at much larger
distances.

\end{abstract}


\section{Introduction}

    Powerful,
highly-collimated,
oppositely directed jets are
observed in
 active galaxies and
quasars (see for example Bridle \& Eilek 1984),
old
compact stars in binaries (Mirabel \& Rodriguez 1994),
 and emission line jets in young stellar
objects (Mundt 1985;  B\"uhrke, Mundt,
\& Ray 1988).
    A broad spectrum
of ideas and models have been put
forward to explain astrophysical jets
(see reviews by Begelman, Blandford,
\& Rees 1984 and Bisnovatyi-Kogan 1993).
    The matter is thought to go to the
jet from the inner region of an
accretion disk surrounding the compact
object - a star or black hole.
The disk matter must then be accelerated
to a velocity higher that the escape
velocity from the central object.
Further, the jet matter should have
sufficient momentum to propagate
through surrounding inter-stellar
or intra-galactic matter out to large
distances.

  An ordered magnetic field
is widely thought
to have an essential
role in jet formation from a
rotating accretion disk.
   Two main regimes have
been considered in theoretical
models, the {\it Poynting flux regime}
where the energy outflow from the
disk is carried mainly by the
electromagnetic field with the
energy carried by the matter
small, and the
{\it hydromagnetic regime} where the
energy is carried by both
the electromagnetic field and
the kinetic flux of matter.
    Poynting flux models for
the origin of  jets
were proposed by
Lovelace (1976) and  Blandford (1976)
and studied further by
 Lovelace et al. (1987)
and Colgate \& Li (1998).
    In these models the
rotation of a Keplerian
accretion disk twists a
poloidal field threading
the disk, and this results
in outflows out of
the disk which carry
angular momentum (in the
twist of the field) and
energy (in the Poynting
flux) away from the disk,
thereby facilitating the
accretion of matter.

    Important  questions to be
answered by the jet models
include the following:
{\bf 1.} What is the main
driving force pushing matter into the jet?
{\bf 2.} What determines the mass flow
rate in the jet ${\dot M_j}$, and what
fraction is this of the accretion rate?
{\bf 3.} What physics determines the
asymptotic speed or Lorentz factor of
the bulk flow? {\bf 4.} What determines
the collimation of the jet and at what
distance from the central object does
the jet become collimated? {\bf 5.}
What is the acceleration mechanism
of leptons to Lorentz factors
$\gamma \sim 10^2 - 10^3$ in the radio jets?
Observations of Blazars indicate that
$\gamma \sim 10^3 - 10^5$ in some objects.

\begin{figure*}[t]
\epsscale{1.4}
\plotone{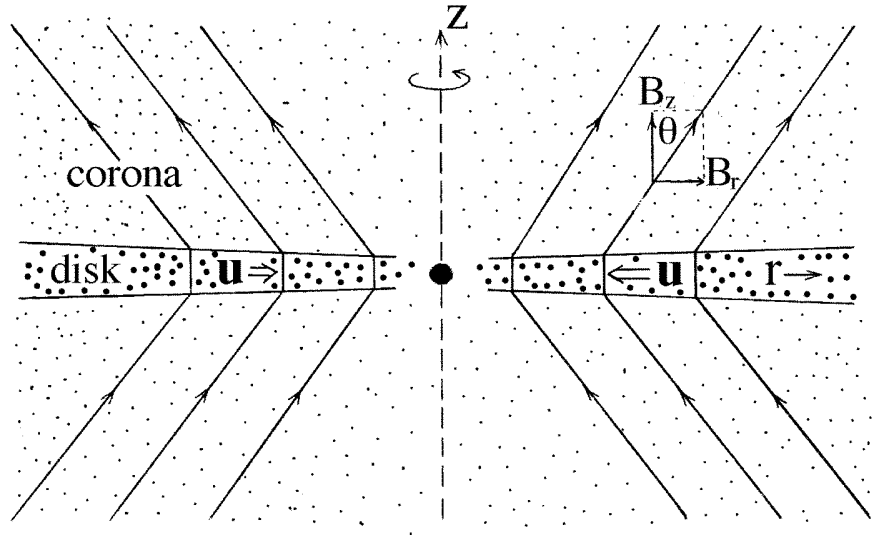}
\caption{Sketch of an accretion disk
threaded by a magnetic field for
conditions which may lead to hydromagnetic
jet formation.}
\label{Fig1}
\end{figure*}

   The focus of recent work
has been the hydromagnetic
regime of jet formation for
the geometry sketched in
Figure 1.
   A strong case for
hydromagnetic
jets as an explanation of
jets in protostellar
systems emerges because the
temperature of the inner
regions of these systems is
insufficient to permit
driving by thermal or
radiation pressure
(K\"onigl \&
Ruden 1993).
   Part of the
investigations have
been analytical or
semi-analytical
and
outgrowths of the
self-similar solution of
Blandford \& Payne (1982)
(Pudritz \&
Norman 1986; K\"onigl 1989;
Pelletier \& Pudritz 1992;
Contopoulos \& Lovelace 1994).
    The outflows in this
model are often referred to
as ``centrifugally driven''
owing to the driving force
close to the disk: If the
poloidal magnetic field
lines diverge from the disk
surface (making an angle
with the $z$-axis of more than
$30^\circ$), then the sum
of the gravitational and
centrifugal forces is in the
$+z$ direction for an MHD
fluid particle which
initially tends to maintain
a constant angular rotation
rate.
   The self-similar models are
unsatisfactory in the respect
that they must be cutoff at
small cylindrical radii,
$r \leq r_{min}$.
  This is the most important region of the
jet flow.
   Observations of optical stellar
jets (Mundt 1985) reveal jet velocities
$\sim 200 - 400 $ km/s, which are
comparable to the Keplerian disk velocity
close to the star's surface.
    This suggests that the jets originate from
the inner region of the disk close to the
star (Shu et al. 1988;
Pringle 1989).

    The limitations of analytical
models
has motivated
efforts to study jet formation using
MHD simulations.
    Simulation
studies of hydromagnetic jet formation
have addressed two main regions:
  The  jet formation region where the
matter enters with  sub
slow-magnetosonic speed and exits
with super fast-magnetosonic speed.
   The second region includes the disk and the
problem of the Velikhov (1959) - Chandrasekhar (1981) -
Balbus-Hawley (1998)
instability
and the resulting 3D MHD
turbulence.
   A number of studies have addressed
the coupled
problem of the disk and near jet
regions (Uchida \& Shibata 1985;
Shibata \& Uchida 1986;
Stone \& Norman 1994;
Bell and Lucek 1995;).
    MHD simulations of the
near jet region have been
carried out by several
groups (Bell 1994; Ustyugova et
al. 1995, 1998;
Koldoba et al. 1995; Romanova et al. 1997, 1998;
Meier et al. 1997; Ouyed \& Pudritz 1997).

   Here, we first review recent results
on  MHD simulations of hydromagnetic
jet formation and later discuss the
Poynting flux regime.
  Section 2
discusses general considerations of
hydromagnetic outflows.
   Sections 3
discusses MHD simulations  which
give non-stationary and stationary
hydromagnetic outflows.
     Sections 4 describes the
Poynting flux regime which
remains to be fully explored by
simulations.
   Section 5
gives the conclusions.
\medskip

\begin{figure*}[t]
\epsscale{1.8}
\plotone{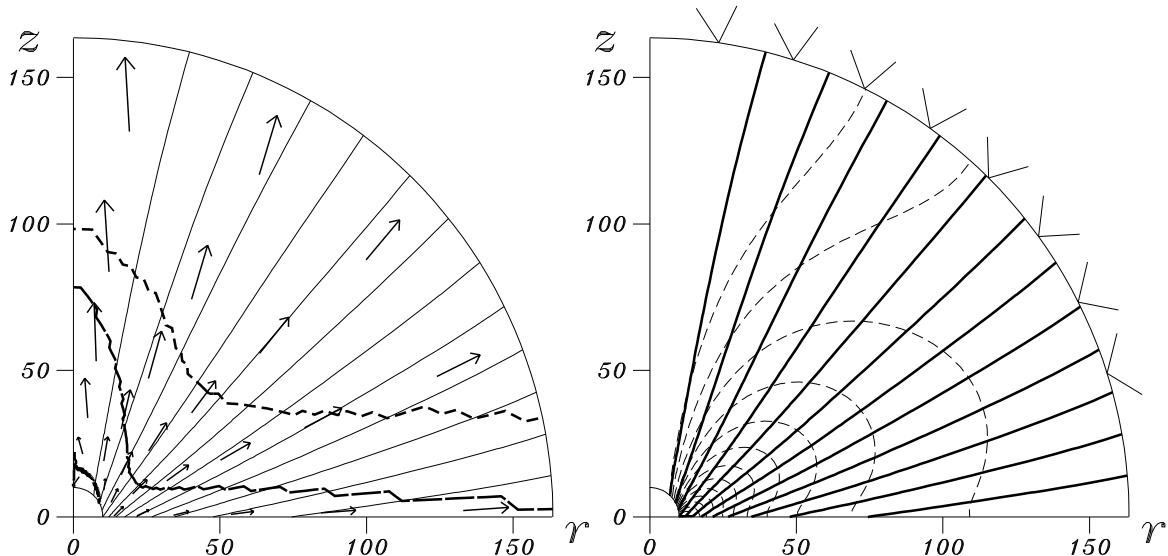}
\caption{
Simulations results for stationary
MHD outflow obtained
using a spherical coordinate system
(Ustyugova et al. 1998).
  The solid lines represent the
poloidal magnetic field,
and the arrows the velocity vectors.
  The dashed lines in the left-hand
plot represent the slow magnetosonic
surface
(the lowest dashed line), the Alfv\'en
surface (the middle line), and the
fast magnetosonic surface (the
top line).
In the right-hand panel, the dashed lines are
surfaces of constant toroidal current density,
while the lines on the outer boundary
are the projections of the fast
magnetosonic Mach cones.}
\label{Fig2}
\end{figure*}

\section{Basics of MHD Outflows
from Disks}

   The main forces which
drive a hydromagnetic outflow from a
disk threaded by a magnetic field
are the centrifugal
force and the magnetic
pressure gradient force.
  If disk has
a hot corona, the pressure gradient may
also be important.
   We neglect the radiative force
but in this regard see Phinney (1987).
   Accreting matter
of the disk carries magnetic field
inward thus generating a $B_r$
component of the magnetic field as
sketched in Figure 1.
   On the
other hand, rotation of
the disk acts to generate a
toroidal component of the field
$B_{\phi}$ ($<0$ if $B_z >0$).

    For a sufficiently inclined
magnetic field
($\theta$ in Figure 1 sufficiently large),
outflows can result from the centrifugal
force (Blandford \& Payne 1982) and/or
the magnetic pressure gradient force
($-{\nabla_z}~B^2_{\phi}/(8\pi)$)
(Lovelace et al. 1989, 1991;
Koupelis \& Van Horn 1989).
 This depends on
the ratio of energy densities at the
base of the outflow at the inner
radius of the disk denoted $r_i$.
Thus, the main parameters are
$
\varepsilon_{th} = (c_s/v_K)_i^2$ and
$\varepsilon_B=(v_{A}/v_K)_i^2$,
where $v_K$ is Keplerian velocity,
$c_s$ the sound speed, $v_A$
the Alfv{\'e}n speed, and the $i$
subscript indicates evaluation on
the surface of the disk at its
inner radius, $r=r_i$.
For $\varepsilon_B \sim 1$
the outflow is magnetically driven, whereas for
$\varepsilon_{th}\sim 1$,
the flow is thermally driven.

   Processes in
the disk are of course coupled to the outflows
(Lovelace et al. 1994, 1997;
Falcke, Malkan, \& Biermann 1995).
   However, it
is difficult to simulate both
regions simultaneously because the
time scales of the accretion
and outflow are in general very
different.
   The accretion is much slower.
   On the other hand the processes
in the disk may involve the small scale
MHD instability of Chandrasekhar,
Velikhov, Balbus, and Hawley,
and therefore
require high spatial resolution.
 Stone \& Norman (1994) attempted
to simultaneously simulate the internal
MHD dynamics of a disk and MHD
dynamics of outflows.
   This proved impractical because
essentially all of the spatial
resolution was needed for treating the
unstable dynamics of the disk.
  Also, there was the problem that
the initial configuration was far
from equilibium.
   Simulation of the internal MHD
disk dynamics has led several groups
to the problem of simulating 3D MHD
turbulence in a sheared flow of a
patch of a disk
(for example, Hawley et al. 1995;
Brandenburg et al. 1995).
   This is a much larger project than that
of understanding MHD outflows.
  At the same time it is widely thought, and
observations of cataclysmic variables
support the view, that the disk
turbulence - including MHD
turbulence - can be modeled approximately using the
 Shakura (1973), Shakura \&
Sunyaev (1973) ``alpha''
viscosity model (Eardley \& Lightman 1975;
Coroniti 1981).
   In contrast with the
internal disk dynamics, there is
theoretical and simulation
evidence that the outflows can
be treated using axisymmetric (2D) MHD
(Blandford \& Payne 1982;
Lovelace et al. 1991; Ustyugova et al. 1995).
  Here, we consider outflows from a disk
represented as a boundary
condition.
    This approach has subsequently
been adopted by other groups
(Meier et al. 1997;
Ouyed \& Pudritz 1997).
   This treatment of the
disk is justified for outflows
from a disk where the accretion speed is small
compared with the Keplerian speed
(Ustyugova et al. 1995).

\section{Numerical Simulations of MHD
Outflows}

   In order to test the
analytical models
of stationary outflows,
MHD simulations of flows
from a disk treated
as a boundary condition have
been carried out by a number
of groups.

\subsection{Non-Stationary Outflows}

   The initial magnetic field
configuration was chosen
so that the magnetic field
was significantly inclined to the disk
($\theta>30^o$) over most of
the disk surface (Ustyugova et al. 1995;
Koldoba et al. 1995).
    The simulations involve solving the
complete system of ideal non-relativistic
MHD equations using a Godunov-type
code assuming axisymmetry but
taking into account all three
components of velocity and magnetic
field.
     Matter of the
corona was initially in thermal
equilibrium with the gravitating center.
  At $t=0$, the disk is set into
rotation with Keplerian velocity
and at the same time matter is pushed
from the disk with a small poloidal velocity
equal to a fraction of the slow magnetosonic
velocity ( $v_p = \alpha v_{sm}$,
with $\alpha=0.1-0.9$).
  A relatively high
temperature and small magnetic field
was considered.
    We found that at the
maximum of the outflow,
matter is accelerated to
speeds in excess of the escape
speed and in excess of the
fast magnetosonic speed within
the simulation region ($\sim 100 r_i$).
   The acceleration is
due to both thermal and magnetic pressure
gradients.
   The outflow
collimates within the simulation
region due to strong amplification,
`wrapping up' of the toroidal
magnetic field and the associated
pinching force.

  However, the outflows are {\it not stationary}.
   The matter flux grows to a peak
and then decreases to relatively small
values.
     The strong collimation of the outflow
reduces the divergence of the field away
from the $z-$axis ($\theta < 30^o$) and this
``turns off'' the outflow of matter and
leads to flow velocities less than the
escape speed.
   Thus, this simulation is
an example of a temporary outburst
of matter to a jet.
    Unfortunately, this type of
flow has a significant dependence
on the initial conditions.

\begin{figure*}[t]
\epsscale{1.5}
\plotone{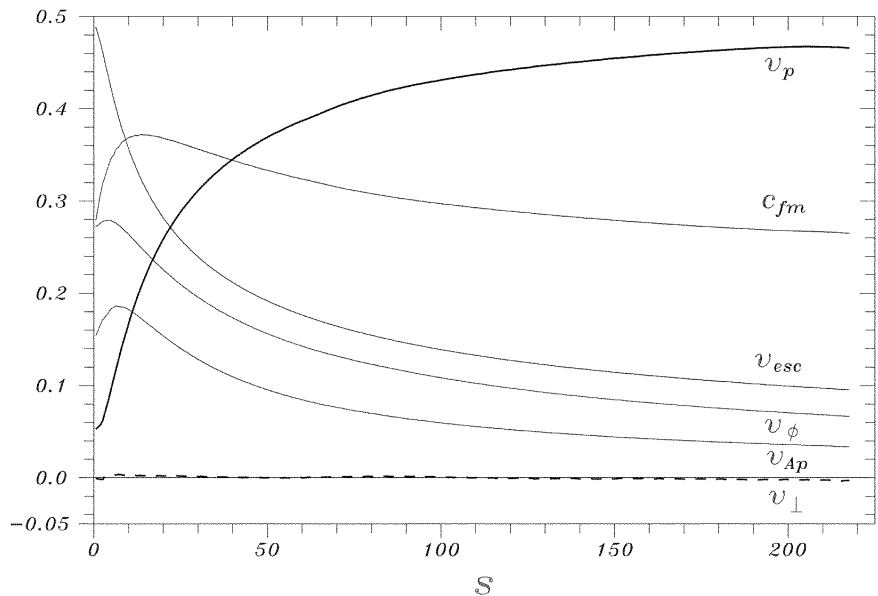}
\caption{
Dependences of different
velocities on distance $s$ measured
in units of $r_i$
from the disk along  the second
 magnetic field line away from
the axis in Figure 2 (Ustyugova
et al. 1998).
  This field line ``starts'' from
the disk  at $r \approx 6r_i$
where it has an angle
 $\theta \approx 28^o$ relative to
the $z-$axis.
   The velocities are measured in
units of $(GM/r_i)^{1/2}$.
   Here, $v_p$ is the poloidal velocity along
the field line and
$v_\perp$ is the poloidal velocity perpendicular
to the field line.
   Also, $v_{Ap}$  is the
poloidal Alfv\'en velocity,  $c_{fm}$
is the fast magnetosonic velocity, and
$v_{esc}$ is the local escape velocity.}
\label{Fig3}
\end{figure*}


\subsection{Stationary Outflows}

 More recently, stationary magnetohydrodynamic
outflows from a rotating
accretion disk have been obtained
by time-dependent axisymmetric
simulations by Romanova et al. (1997)
and systematically analyzed by
Ustyugova et al. (1998).
 The initial magnetic field in the
latter work was taken to
be a split-monopole poloidal
field configuration (Sakurai 1987)
frozen into the
disk.
   The disk was treated as a perfectly
conducting, time-independent density
 boundary [$\rho(r)$]
in Keplerian rotation which is different
from our earlier specification of a
small velocity outflow (\S 3.1, Ustyugova
et al. 1995).
  The outflow velocity
from the disk is
determined self-consistently from the MHD
equations.
   The temperature of
the matter outflowing from
the disk is  small in the
region  where the magnetic
field is inclined away from
the symmetry axis
($c_s^2 \ll v_K^2$), but
relatively high
($c_s^2 ~{\buildrel < \over \sim}~v_K^2$)
at very small radii in the disk
where the magnetic field is not inclined
away from the axis.
   We have found a large class of stationary
MHD winds.
   Within the simulation region, the outflow
accelerates from thermal velocity ($\sim c_s$)
to a much larger asymptotic poloidal flow
velocity of the order of  $0.5\sqrt{GM/r_i}$,
where $M$ is the mass of the central object
and $r_i$ is the inner radius of the disk.
   This asymptotic velocity is much larger than
the local escape speed and is larger than
fast magnetosonic speed by a factor of $\sim 1.75$.
  The {\it acceleration distance} for the outflow, over
which the flow accelerates from $\sim 0$
to, say, $90\%$ of the asymptotic speed, occurs
at a flow distance of about $80 r_i$.
   The flows are approximately {\it spherical
outflows}, with
only small collimation within the simulation region.
   The {\it collimation distance} over which
the flow becomes collimated (with divergence less
than, say, $10^o$) is {\it much larger} than the size of
our simulation region.
   Close to the disk the outflow is driven by
the centrifugal force while at all larger
distances the flow is driven by the
magnetic force which is proportional to $-{\bf \nabla}
(rB_\phi)^2$, where $B_\phi$ is the toroidal field.

  The stationary MHD flow solutions allow us
(1) to compare the results  with MHD theory
of stationary flows,
(2) to investigate the influence of different
outer boundary conditions
on the flows, and (3) to investigate the influence
of the shape of the simulation region
 on the flows.
   The ideal MHD integrals of motion (constants
on flux surfaces discussed by Lovelace et
al. 1986) were
calculated along magnetic field
lines and were shown
to be constants with accuracy $5-15 \%$.
   Other characteristics of the
numerical solutions were compared with the theory,
including conditions at the
Alfv\'en surface.

\begin{figure*}[b]
\epsscale{1.4}
\plotone{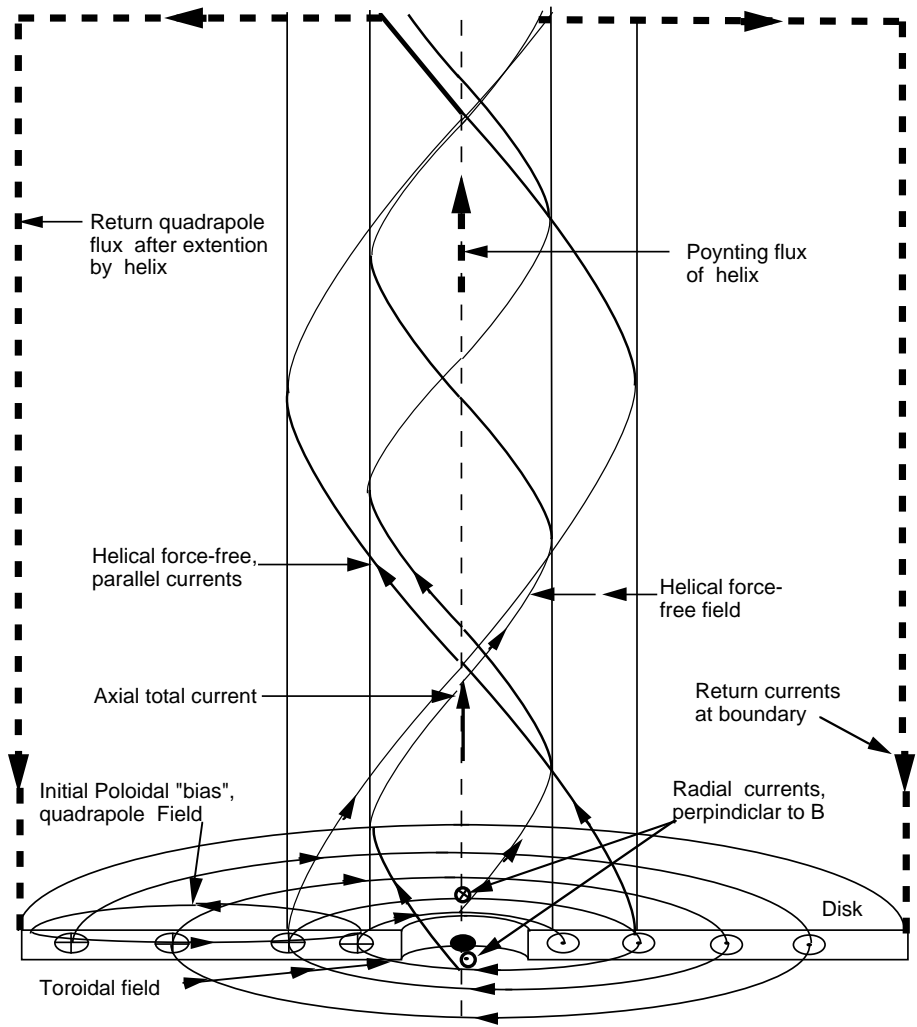}
\caption{Field configuration of
a Poynting flux jet from Colgate
and Li (1998).}
\label{Fig4}
\end{figure*}

    Different outer boundary conditions
on the  toroidal component
of the magnetic field can significantly
influence the calculated flows.
    The commonly used ``free'' boundary condition on
the toroidal field leads
to  artificial magnetic
forces on the outer boundaries,
which can give spurious collimation
of the flows.
   New outer boundary conditions
which do not give artificial forces have
been proposed and
investigated by Ustyugova et al. (1998).

   The simulated flows
may also depend  on the shape
of the simulation region.
   Namely, if the simulation region
is elongated in the $z-$direction,
then Mach cones on the outer
cylindrical boundaries may be
partially directed inside the
simulation region.
   Because of this, the boundary can have
an artificial influence on the calculated
flow.
   This effect is reduced if the
computational region is approximately square
or if it is spherical as in Figure 2.
  Simulations of MHD outflows with an
elongated computational
region can lead to
{\it artificial} collimation
of the flow.

   Recent simulation studies have treated
MHD outflows from disks with
more general initial ${\bf B}$
field configurations, for example, that
where the poloidal field has different
polarities as a function of radius
(Hayashi, Shibata, \& Matsumoto 1996;
Goodson, Winglee,
\& B\"ohm 1997;
Romanova et al. 1998).
   The differential rotation
of the foot-points of
${\bf B}$ field loops at different
radii on the
disk surface causes
twisting of the coronal magnetic
field, an increase in the
coronal magnetic energy, and an
opening of the loops in
the region where the magnetic pressure
is larger than the matter pressure
($\beta {\buildrel < \over \sim} 1$)
(Romanova et al. 1998).
   In the region where
$\beta {\buildrel >\over \sim}1$,
the loops may be only
partially opened.
  Current layers form in the
narrow regions
which separate oppositely
directed magnetic field.
    Reconnection occurs in these
layers as a result of the
small numerical magnetic
diffusivity.
    In contrast with the case of the
solar corona,
there can be a steady outflow of energy and
matter from the disk surface.
 We find that the power output
mainly in the form of a
Poynting flux.
    Opening of magnetic field loops
and subsequent closing
can give
reconnection events which may be responsible
for X-ray flares in disks
around both stellar mass objects
and massive black holes (Hayashi et al. 1996;
Goodson, Winglee,
\& B\"ohm 1997;
 Romanova et al. 1998).

\section{Poynting Flux Jets}

   In a very different regime from
the hydromagnetic flows discussed
in \S 3,  a Poynting flux jet
transports energy and angular
momentum mainly by the
electromagnetic fields with
only small contributions of the
matter (Lovelace et al. 1987;
Colgate and Li 1998).
  A steady Poynting jet - sketched
in Figure 4 - can be
characterized in the lab frame
by its asymptotic (large distance)
magnetic field $B_\phi = -B_0 [r_o/r_j(z)]$
at the jet's edge, $r=r_j(z)$, where
$r_0$ is the jet's radius at $z=0$
and $B_0$ is the poloidal field strength
at this location.
   The electric field in the
jet is ${\bf E} = -{\bf v}\times {\bf B}/c$
and consequently the energy flux
or luminosity of the jet is
$L_j = v B_0^2r_0^2/8 \sim 4\times 10^{45}
{\rm erg/s}(v/c)(r_0/10^{14}{\rm cm})^2
(B_0/10^4 {\rm G})^2$.
  Propagating disturbances in such field
dominated
jets provide a simple, but
self-consistent physical
model for the gamma ray flares observed
in Blazars (Romanova \& Lovelace 1997;
Levinson 1998;
Romanova these proceedings).
   Owing to pair production close to
the black hole, the main constituent of
a Poynting flux jet may be electron-positron
pairs.

\section{Conclusions}

    MHD simulations carried
out by a number of groups
over the last several years
support the idea that an
ordered magnetic field of
an accretion disk can give
powerful outflows of matter,
energy, and angular momentum.
   The studies so far have
been in the hydromagnetic
regime and find
asymptotic flow speeds of the
order of the maximum Keplerian
velocity of the disk.
    In contrast,
observed VLBI jets in quasars and
active galaxies point to bulk
Lorentz factors of order $10$ - much
larger than the disk Lorentz factor.
   This may be a result of the
relativistic dynamics not
included here, but more likely
it reflects the fact that these
jets are in the Poynting flux
regime.
   Also, these jets  may involve
energy extraction
from a rotating black hole (Blandford
and Znajek 1977;  Livio et al. 1998).

\acknowledgments

  It is a pleasure to thank
the organizers for a most memorable
meeting.
This work was supported in part by
NSF grant AST-9320068 and NASA
grant NAG5 6311.
   The Russian
authors were supported in part
by RFFI Grant 96-02-17113.
   Also, the research described here
was made possible in part by
Grant No. RP1-173 of the U.S.
Civilian R\&D Foundation for the Independent
States of the Former Soviet Union.
\medskip

\noindent{\bf Discussion}
\medskip

\noindent{\it Duccio Macchetto:}
What is the collimation distance in
the case of relativistic jets?
  Does it also scale and collimate
at distances $\sim 100 r_i$?
  If so it would still be too
close to be resolved spatially.

\medskip
\noindent{\it Richard Lovelace:}
  Simulations of the origin of
relativistic jets remains to
be done systematically.
   In the hydromagnetic regime,
the flows are likely to be similar to
those in the
non-relativistic limit with
little collimation at distances
$ < 100r_i$.
   In contrast I believe that
the Poynting flux jets
collimate rapidly over distances
of order several $r_i$.

\medskip
\noindent{\it Julian Krolik:}
What parameters determine the mass
outflow rate?

\medskip
\noindent{\it Richard Lovelace:}
   For the hydromagnetic outflows
we find in
the non-relativistic limit  that
roughly  $\dot{M}_j \propto r_i^2 B_i^2$,
where $r_i$ is the inner radius of
the disk [$ (2-6)GM/c^2$] and $B_i$
is the poloidal field at $r_i$.
  However, $\dot{M}_j$ will be
less than some fraction of the
disk accretion rate.
  For the Poynting flux jets, $\dot{M}_j$
is negligible compared with the
disk accretion rate.

\medskip
\noindent{\it Michael Dopita:}
In the wind dominated case, can
the wind itself become of sufficiently
high column density to become optically
thick to Thompson scattering and thereby
form a photosphere in the wind?

\medskip
\noindent{\it Richard Lovelace:}
  A hydromagnetic outflow with $\dot{M}_j$
say one-tenth of the disk accretion
rate would have a photosphere  well
above the disk.

\end{document}